# Revealing electrically undetectable room temperature surface-mobility of bulky topological insulators by spectroscopic techniques

**Authors**

Bumjoo Lee[1,2,*], Jinsu Kim[3], Jonghyeon Kim[4], Na Hyun Jo[3,#], Yukiaki Ishida[5], So Yeun Kim[1,2], Min-Cheol Lee[1,2], Inho Kwak[1,2], Shik Shin[5], Kyungwan Kim[6], Jae Hoon Kim[4], Myung-Hwa Jung[3], Tae Won Noh[1,2], and Byung Cheol Park[1,2,†]

**Affiliation**

[1]Center for Correlated Electron Systems, Institute for Basic Science, Seoul 08826, Republic of Korea

[2]Department of Physics and Astronomy, Seoul National University, Seoul 08826, Republic of Korea

[3]Department of Physics, Sogang University, Seoul 04107, Republic of Korea

[4]Department of Physics, Yonsei University, Seoul 03722, Republic of Korea

[5]ISSP, University of Tokyo, Kashiwa, Chiba 277-8581, Japan

[6]Department of Physics, Chungbuk National University, Cheongju 28644, Republic of Korea

*First author (B. L, kimphysics@snu.ac.kr)

†Corresponding author (B. C. P, topologicalmeta@gmail.com)

#Current affiliation (N. H. Jo): Ames Laboratory, Iowa State University, Ames, Iowa 50011, USA

## Abstract

High surface-mobility, which is attributable to topological protection, is a trademark of three-dimensional topological insulators (3DTIs). Exploiting surface-mobility indicates successful application of topological properties for practical purposes. However, the detection of the surface-mobility has been hindered by the inevitable bulk conduction. Even in the case of high-quality crystals, the bulk state forms the dominant channel of the electrical current. Therefore, with electrical transport measurement, the surface-mobility can be resolved only below-micrometer-thick crystals. The evaluation of the surface-mobility becomes more challenging at higher temperatures, where phonons can play a role. Here, using spectroscopic techniques, we successfully evaluated the surface-mobility of $Bi_2Te_3$ (BT) at room temperature (RT). We acquired the effective masses and mean scattering times for both the surface and bulk states using angle-resolved photoemission and terahertz time-domain spectroscopy. We revealed a record-high surface-mobility for BT, exceeding 33,000 $cm^2 V^{-1} s^{-1}$ per surface sheet, despite intrinsic limitations by the coexisting bulk state as well as phonons at RT. Our findings partially support the interesting conclusion that the topological protection persists at RT. Our approach could be applicable to other topological materials possessing multiband structures near the Fermi level.

**Introduction**

Three-dimensional topological insulators (3DTIs) have attracted much attention as novel quantum states of matter hosting helical Dirac fermions[1-3]. 3DTIs are topologically distinct from normal insulators, due to their strong spin-orbit interaction (SOI) and time-reversal symmetry (TRS)[1-3]. Spin splitting and band inversion occur due to the SOI, resulting in spin($s$)-polarized surface states[1-4]. Due to the TRS, the $s$-polarized surface states cross at the time-reversal (TR) invariant momentum[1-3]. This enables the formation of a linear energy ($E$)–momentum ($k$) dispersion[1-6] called a Dirac cone, as in graphene. Distinct from graphene, the linear surface state hosts spin helical Dirac fermions, whose $s$ is locked perpendicular to the $k$[4,7]. The surface state is called topological surface state (TSS), as its $s$-$k$ locking is associated with the nontrivial topology of 3DTIs[1-3]. $Bi_2Te_3$ (BT) and $Bi_2Se_3$ (BS) are prototypical binary 3DTI materials, with a bulk band gap of a few hundred meV and an in-gap TSS[5,8].

Interestingly, the mobility of the TSS can exceed those of 3D noble metals or doped semiconductors. Mobility is a measure of how quickly charge carriers can move when an electric field is applied[9]. For Fermi liquids, including Dirac fermions, the mobility is defined as $\mu \equiv e\tau/m^*$ [ref. 9]. Here, $e$ is the elementary charge, $\tau$ is the scattering time, and $m^*$ is the effective mass. In a Dirac state, the effective ('fictitious' relativistic[10]) mass is determined by $m^* = (E_F - E_D)/v_F^2$, where $E_F$ is the Fermi level, $E_D$ is the Dirac point energy, and $v_F$ is the Fermi velocity[11]. For the TSS, the $m^*$ is generally small ($\lesssim 0.1\ m_e$)[12,13], and even converges to zero at the Dirac point ($E_F = E_D$). Also, the $\tau$ of the TSS can be very long if the TRS is conserved, because its $s$-$k$ locking blocks backscattering of the charge carriers by impurities[7]. Based on its small $m^*$ and long $\tau$, the TSS mobility can be estimated to be very high. This suggests its potential for various applications[14,15]

However, it is challenging to observe such high TSS mobility of bulky 3DTI crystals, using electrical transport measurements[8,16-30]. The bulk conduction, arising from unintentional doping[31], also contributes to the transport characteristics as the TSS conductance[32]. Although the TSS is a more conducting channel than the bulk state, an electrical current can mostly flow through the bulk channel in the case of thick 3DTIs. Then, the transport measurement provides the bulk mobility as a dominant signal. Some trials have been performed to obtain the TSS mobility, e.g., Bansal et al.[33] performed transport measurements while applying a magnetic field. They measured the Hall (transverse) conductance and analyzed it with a two-carrier model, to distinguish the TSS and bulk mobilities. We noted that those trials were, in fact, quite suitable for thin 3DTIs, below a micrometer, where the TSS and bulk conductance were comparable[33]. However, in the case of thick 3DTI samples, the application of two-carrier analysis would be challenging due to the dominant bulk conductance.

Although intrinsically limited by phonons at room temperature (RT)[34], the TSS mobility is expected to be high due to the *s-k* locking at RT, for which there is both theoretical[35] and experimental[4,36] evidence. However, such high TSS mobility has never been experimentally observed at RT, which impedes the practical applications of 3DTIs. This led us to doubt the applicability of the results of previous transport studies, so we attempted to design an appropriate method for addressing this issue.

In this study, by adopting spectroscopic techniques, we successfully revealed the ultrahigh TSS mobility of bulky 3DTI BT at RT. These methods allowed us to observe the excitation of electronic states, not the electrical current. Therefore, they provided robust information on the effective mass and/or scattering time, regardless of the extrinsic factors, such as the thickness of the sample, in sharp contrast to the transport measurement. Laser ARPES was used to determine the effective mass of charge carriers from the high-resolution band dispersion. Terahertz time-domain spectroscopy (THz-TDS) was employed to provide the scattering time of the charge carriers from the broadband spectrum. Their combination allows us to obtain ultrahigh TSS mobility at RT in a non-contact and non-destructive manner. We propose a new approach for measuring the TSS mobility at RT, which has previously been challenging, even in the case of high-quality 3DTI samples[31,32].

**Growth and characterization of $Bi_2Te_3$ (BT) single crystal**

In this section, we describe how we grew the BT crystal, and how to confirm the quality of the sample with various characterization tools.

Using the Bridgeman method[24], we prepared high-quality BT single crystals. The raw material, a mixture of Bi and Te powders, was melted at a soaking temperature of 800°C for 16 h, slowly cooled to 550°C, kept for 3 days at 550°C, and then finally cooled to RT. Additional Te (2%) was added to the stoichiometric raw material to compensate for the loss of Te, which is highly volatile at high temperatures. The as-prepared single crystals were cleaved well, as indicated by the shiny surface of the exposed basal plane.

To verify the good crystallinity of our BT samples, we performed X-ray diffraction (XRD), scanning tunneling microscopy (STM), and transmission electron microscopy (TEM) measurements. In Fig. 1a, XRD patterns of the cleaved surface show a family of (003) peaks, representing the rhombohedral (*R*-3*m*) symmetry of BT and yielding the *c*-axis lattice constant of 30.47 Å. Each peak position was in good agreement with that in the BT reference (red dashed line, ICSD code. 15753), without any impurity peaks. In Fig. 1b, the atomic-resolution STM image exhibited a hexagonal unit cell of Te atoms in the top-most layer; the corresponding spacing of Te atoms was 4.32 Å. Aberration-corrected scanning TEM images revealed that the quintuple layers (Te-Bi-Te-Bi-Te) of BT were clearly separated by a van der

Waals gap, without any lattice deformation. These results confirm the high crystallinity of the BT single crystals used in this study.

To characterize the electronic properties, we further carried out temperature ($T$)-dependent transport measurements (see **Supplementary S1**). The transport data show the oscillatory signal of the electrical resistivity as a function of $T$. According to a previous report[17], this indicates that the $E_F$ of our BT sample lies near the bulk conduction band (BCB).

**Acquisition of effective mass with (time-resolved) ARPES**

In this section, we present our strategy for obtaining the effective masses of charge carriers with ARPES. We employed a laser-based ARPES system (see ***Methods***) that provides the *E-**k*** dispersion of the electron-occupied TSS with a high energy resolution of ~15 meV. We also carried out time-resolved ARPES to reveal the BCB dispersion above the $E_F$ by applying an optical pump beam. Note that we performed all the ARPES measurements at RT on *ex-situ* grown and *in-situ* cleaved BT sample.

Figure 2a shows the equilibrium ARPES data, including the well-defined band structure of our BT sample. The TSS with a linear *E-**k*** dispersion is clearly present over the region from the $E_D$ to the $E_F$. However, the BCB is barely visible in the ARPES data because of the low $E_F$. This indicates that our BT sample has negligible band bending, which arises from the unintentional doping by impurities[31]. To better expose the BCB lying above the $E_F$, we integrated the ARPES signal over the emission angle $\theta$ range of $-1° \leq \theta \leq 1°$ (around $-0.01 \text{ Å}^{-1} \leq k \leq 0.01 \text{ Å}^{-1}$). This resulted in a peak around $E - E_F \cong 0.1 \text{ eV}$ corresponding to the BCB, as shown in Fig. 2b.

From the ARPES data, we confirmed that the TSS of our BT sample can be described by Dirac theory. The ARPES data indicated that the $E_F$ value was 167 meV above $E_D$, yielded a Fermi velocity, $v_F$, of $3.4 \times 10^5$ m/s, and a Fermi wavenumber, $k_F$, of 0.075 $\text{Å}^{-1}$. [We note that $E_F$ corresponds to the very low impurity density of approximately $3 \times 10^{14}/\text{cm}^3 - 10^{17}/\text{cm}^3$ [ref. 31].] The ARPES data support our assumption that the TSS of our BT sample falls into the regime where the Dirac-conic description is valid. Therefore, the charge carriers in the TSS can be described as Dirac quasiparticles, which are distinct from the conventional quasiparticles in quadratic bands[37].

We then calculated the effective mass of the TSS carriers, $m^*_{\text{TSS}} \cong 0.25 \ m_e$, from the linear *E-**k*** dispersion of the TSS in the ARPES data. To calculate the effective mass, we applied the semiclassical equation typically used for Dirac materials[38]: $m^* = \hbar k_F / v_F$, where $\hbar$ is the reduced Planck constant.

Using the $k_F$ and $v_F$ values above, we finally obtained the effective mass of the TSS carriers, $m^*_{TSS} \cong 0.25~m_e$. The value is in the reasonable boundary according to the literature[17,24].

Next, from the quadratic BCB in the time-resolved ARPES data in Fig. 2c, we acquired the effective mass of the bulk carriers, $m^*_{BCB} \cong 0.25~m_e$. In this case, we used the optical pump beam to reveal the BCB in the nonequilibrium state. Otherwise, the dispersion of the BCB is hidden above the $E_F$, as in the equilibrium data. Figure 2c shows the time-resolved ARPES data, which show the quadratic dispersion of the BCB described by $E = \hbar^2 k^2/(2m^*)$. Through the model fitting (white curves in Fig. 2c), we obtained the effective mass of the bulk carriers, $m^*_{BCB} \cong 0.25~m_e$. To compare with the equilibrium data, we presented the integrated nonequilibrium ARPES intensity in Fig. 2d. These results show how, and how many, electrons were promoted by the optical pumping.

Interestingly, we observed that $m^*_{TSS}$ is very similar to $m^*_{BCB}$, indicating the importance of the scattering time for understanding the mobility. As mentioned above, $m^*_{TSS}$ was varied while $m^*_{BCB}$ was held constant as $E_F$ was changed. Therefore, at a certain $E_F$, both effective masses are equivalent. According to our ARPES data, this unique phenomenon occurs in our BT sample with an $E_F$ of ~167 meV. This observation is interesting by itself, given that their band dispersions are evidently distinct. Also, this implies that the effective mass is not crucial for distinguishing the mobilities of the TSS and bulk carriers. This naturally requires further investigation of the scattering time.

**Scattering time measurement utilizing THz-TDS**

In this section, we describe our method for calculating the scattering times of the charge carriers with THz-TDS. According to our ARPES data (Fig. 2a), the TSS crosses at $E_F$, while the BCB nearly touches it. In this case, two types of the charge carriers, i.e., both the TSS and bulk carriers, should appear in the low-energy electrodynamic response at RT[39].

The low-energy optical conductivity $\tilde{\sigma}(\omega)$ provides the scattering time of the free charge carriers. The terahertz (THz) electric field enables the exploration of low-energy (1 THz = 4.1 meV) excitation of the free carriers[40]. The free carriers' $\tilde{\sigma}(\omega)$ can be modeled by Drude theory[9,39], as long as they are weakly interacting. The alternating current (AC) Drude conductivity is represented by[9,39]:

$$\tilde{\sigma}_{Drude}(\omega) \equiv \sigma_{1,Drude} + i\sigma_{2,Drude} = \frac{\sigma_0}{1-i\omega\tau} = \frac{\pi}{15}\frac{1}{4\pi}\frac{\tau\omega_p^2}{1-i\omega\tau},$$

where $\sigma_0$ is the direct current (DC) conductivity, $\omega_p$ is the plasma frequency, and $\tau$ is the scattering time. The scattering time is directly related to the width of the Drude peak, namely, $\tau = (HWHM)^{-1}$, where HWHM is the half width at half maximum (HWHM) of the Drude peak in the frequency domain.

We used the THz $\tilde{\sigma}(\omega)$ to acquire the scattering times of the charge carriers in our BT sample. Here, we exploited the strength of THz spectroscopy, i.e., the separability of the scattering times. This means that different types of free carriers exhibit different scattering times in the $\tilde{\sigma}(\omega)$. In fact, the $\tilde{\sigma}(\omega)$ has often been used to distinguish the scattering times of charge carriers originating from different electronic states[41-43]. Here, we applied this to our BT sample, where two different scattering times should appear in the THz $\tilde{\sigma}(\omega)$.

We carried out transmission THz spectroscopy (see **Methods** for more details) of a cleaved BT single crystal. [Note that the transmission measurement usually provides a more reliable scattering time than the reflection one]. For the transmission measurement, we cleaved the BT single crystal down to 6 μm by mechanical exfoliation. Then we mounted the cleaved crystal on a sample holder with a hole (~3-mm diameter). We repeatedly measured transmitted THz pulses through the holder (reference signal) and the cleaved BT crystal on the holder (sample signal) in the time domain.

We then obtained the complex transmittance (see **Supplementary S2**) through fast Fourier transform of the transmitted time-domain THz pulses. Figure 3a and 3b, respectively, show the real (squares) and imaginary (circles) parts of the measured transmittance. As in the previous report[44], the transmittance spectra can be explained in terms of three absorptions: two Drude peaks associated with the TSS or bulk states at $E_F$ and a Fano-type oscillator reflecting the infrared-active phonon with a center frequency at 1.7 THz.

To acquire the optical conductivity, we fitted a model to the transmittance. We introduced the initial model conductivity using two Drude terms and the aforementioned Fano oscillators. To reduce the number of free parameters, we used an important constraint: the plasma frequency of the TSS Drude term was fixed to match ARPES and Fourier transform infrared (FTIR) spectroscopy measurements (see **Supplementary S3**). Then, the model conductivity was converted into the model transmittance. The modeled transmittance was repeatedly constructed and fitted to the experimental transmittance until the residual was minimized (see **Supplementary S4**). The best fitting results are indicated by red lines in Fig. 3a and 3b.

Note that, especially for more accurate $\tilde{\sigma}(\omega)$ calculation, we designed a nonuniform structure of 3DTI, as shown in Fig. 3c. Our model reflects the fact that the TSS is not perfectly localized on the surface but penetrates the bulk with an exponentially decaying tail[45]. In this inhomogeneous model, we set the spatial dispersion, $\delta$, of TSS to 2.5 nm. We judged this to be reasonable by comparing it to previous reports[46-48]. Then, to properly link the model to the $\tilde{\sigma}(\omega)$, we used a variational method based on a transfer matrix method (see **Methods**). The bulk was assumed to be uniform over the entire sample of thickness $d_{\text{THz}} \sim 6$ μm. We confirmed that the nonuniform model provides a better goodness of fit than the uniform model (see **Supplementary S5**).

We acquired $\sigma(\omega)$ separately for the TSS, $\sigma$ (TSS), and the bulk, $\sigma$ (bulk). Figure 3d shows the real parts of their conductivities (the complex conductivities are presented in **Supplementary S4**). Through the fitting, we obtained two distinct sharp (blue line) and broad (orange line) Drude peaks, as shown in Fig. 3d. We then assigned the sharp Drude peak to the response of the TSS carriers. Otherwise, the plasma frequency of the TSS carriers is highly inconsistent with that obtained from ARPES and FTIR data (see **Supplementary S6**). The TSS Drude peak corresponds to the downturn in the real transmittance (or real dielectric function) below 3 meV. Those low-energy features are therefore the signatures of the TSS (see **Supplementary S7**). Subsequently, the broad Drude peak can be attributed to the response of the bulk carriers.

We found that the TSS carriers are highly metallic and scatter little. As shown in Fig. 3d, the TSS Drude peak exhibits DC conductivity, $\sigma_{\text{DC}}$ (TSS) $\cong 47{,}000\ \Omega^{-1}\cdot\text{cm}^{-1}$, representing the super-metallic behavior of Dirac fermions[49]. Also, the scattering time of the TSS Drude peak was $\tau$ (TSS) $\cong 4.8$ ps. Interestingly, this implies that the TSS carriers move coherently without any relaxation for a long time, of 4.8 ps. We note that the scattering time, as a key physical quantity in this study, was carefully determined (see **Supplementary S8**). From the scattering rate, we also deduced the mean free path of the TSS, $l$ (TSS) $= v_F \tau \cong 1.63$ μm, which is comparable to or greater than the reported transport values[16-19,25,26].

On the other hand, the bulk carriers are bad metallic and highly scattered in comparison with the TSS carriers. As shown in Fig. 3d, our fitting results provided the DC conductivity of the bulk carriers, $\sigma_{\text{DC}}$ (bulk) $\cong 170\ \Omega^{-1}\text{cm}^{-1}$. The value of the $\sigma_{\text{DC}}$ (bulk) was two orders of magnitude smaller than that of the $\sigma_{\text{DC}}$ (TSS). This indicates that the bulk states are not good metallic channels. Furthermore, we deduced that the scattering time of the bulk carriers is $\tau$ (bulk) $\cong 0.16$ ps. This value is 30 times less than that of the $\tau$ (TSS). Both the DC conductivity and scattering time are similar to the previously reported values for the bulk-dominant TIs (see **Supplementary S9**).

We examined whether or not our THz results were consistent with the transport data at $T = 300$ K. At 300 K, the DC limit of the $\sigma_{\text{DC}}$ (bulk) from our THz-TDS was approximately 170 $\Omega^{-1}\text{cm}^{-1}$ (Fig. 3d) whereas the DC conductivity from the transport experiment was approximately 413 $\Omega^{-1}\text{cm}^{-1}$ (**Supplementary S1**). Thus, they were at least consistent in the order of magnitude. This probably implies that the transport measurement mostly detects the electrical current of the bulk carriers. The difference in these conductivities may originate from the possible error (± 1 μm at maximum) in the thickness used for the bulk THz conductivity calculation or the nonuniformity (roughly ~2 μm) of the exfoliated crystal. Taking into account these factors, the bulk THz conductivity can reach 340 $\Omega^{-1}\text{cm}^{-1}$ at maximum, which is fairly close to the DC transport conductivity. However, we note that these errors have a negligible effect on the TSS THz conductivity value.

**Mobilities of the TSS and bulk state at RT**

Using a combination of ARPES and THz spectroscopy, we ultimately obtained the TSS mobility at RT. To calculate the TSS mobility, we used the formula $\mu\,(\text{TSS}) \equiv ev_F\tau/\hbar k_F$ as applied for Dirac materials, such as graphene[39]. We used $v_F \cong 3.4 \times 10^5$ m/s and $k_F \cong 0.075$ Å$^{-1}$ obtained from our ARPES data, in addition to $\tau\,(\text{TSS}) \cong 4.8$ ps from our THz data. This calculation provided the TSS mobility of 33,000 cm$^2$V$^{-1}$s$^{-1}$ per TSS at RT. Considering the possible error due to Fermi surface warping (about a 5% variation in $k_F$), the TSS mobility varied over the range 32,000–34,000 cm$^2$V$^{-1}$s$^{-1}$. Such a high $\mu\,(\text{TSS})$ is surprising, given that $\mu \propto \sigma \propto T^{-1}$ due to the sizable electron-phonon coupling at RT (above the Debye temperature, $T_{\text{Debye}}$ = 145 K [ref. 50]).

Our TSS mobility at RT is the highest mobility among reported Bi-based TIs[30,51]. We plotted the mobility values of 3DTIs for comparison in Fig. 4. These values were collected from the published reports[8,16-30]. The mobility values were diverse, over 400–29,000 cm$^2$V$^{-1}$s$^{-1}$, even at low temperatures (LTs)[8,16-30]. Notably, our RT $\mu\,(\text{TSS})$ is even higher than the LT mobility values in the literature. This observation is somewhat contradictory to conventional cases where at higher temperatures, especially above the $T_{\text{Debye}}$, phonons act as additional scattering centers for propagating electrons[9]. Also, our sample's mobility is greater than cases where the doping level was tuned closely to the Dirac point[16-19,25,26], which does not accord with our expectation.

Here we provide a reasonable explanation for the various mobilities of 3DTIs shown in Fig. 4. To construct a unified picture covering all references, we considered three main factors: phonons, bulk state hybridization, and impurities that impede the transport properties of the TSS. The bulk state can play a certain role in the 3DTIs, in contrast to graphene, in which the Dirac fermion mobility is primarily limited by phonons at RT[52,53]. [We ignored the limitation of the mobility imposed by the substrate.] In the bulk-dominant regime where the Fermi level is greater than our case, TSS-bulk coupling can occur relatively easily[54-57]. This speculation is reasonably supported by the fact that the bulk Drude peak of our BT sample is similar to Drude peaks in highly doped 3DTIs[55,58,59] (see **Supplementary S9**). With the role of the bulk state, it is possible that the LT mobility in the highly doped regime may be lower than that at RT in the lowly doped regime[21,60]. Additionally, the TSS mobility relies on the charged impurity density directly related to the $E_F$, as is common in semimetals[20,31]. In binary compounds, only one type of charged impurity predominantly determines the Fermi level (e.g., Te defects for BT and Se defects for BS). In complex compounds such as $(Bi_xSb_{1-x})_2(Te_ySe_{1-y})_3$, additional dopants can introduce more impurities and complicate the band structure[16,61]. This could be the origin for the lower mobilities of the lowly doped compounds than binary compounds like ours[20].

How can we explain the RT mobility difference between the TSS and bulk state in our BT sample? In this work, we found the mobility of the bulk state, $\mu$ (bulk) $\cong 1{,}160~\text{cm}^2\text{V}^{-1}\text{s}^{-1}$, which is almost 30 times less than $\mu$ (TSS). There are two possible reasons for such a difference: effective mass and/or scattering time. Given that the masses of the two carriers are similar (Fig. 2a and 2c), the scattering time can be the dominant factor determining their mobilities. The scattering times could differ, given that the TSS is quasi-2D (mostly confined within 2.5 nm[46-48]), whereas the bulk states are 3D. However, our observation that $\tau$ (TSS) $\gg \tau$(bulk) is inconsistent with the conventional view regarding shorter scattering times at lower dimensions (where electron-electron scattering is increasing, whereas electron-phonon scattering is more or less the same[62].) Therefore, we conclude that the TSS and bulk state scattering times are different primarily due to *s-k* locking, as described in the literature[7,44,53,58,63,64].

**Discussion**

In this section, we discuss the generality of our findings, the appropriateness of our analyses, the importance of our work and its potential implications.

To allow comparison with our results, let us discuss why the transport measurement is complicated, particularly for the study of 3DTI; the TSS mobility can only be obtained in the thin limit. According to Ohm's law, $V = I/G$, where V is the voltage, $I$ is the current, and $G$ is the conductance. As mentioned previously, 3DTIs have heterostructures (Fig. 3c) composed of TSS and a bulk state. Therefore, the conductance of 3DTIs can be written as $G = [\sigma(\text{TSS}) \cdot \delta + \sigma(\text{bulk}) \cdot d_{\text{trans}}](w/l)$. Here, $w$ is width of the sample, and $l$ is the distance between the applied voltages $V+$ and $V-$. Plugging in $\sigma$ (TSS), $\sigma$ (bulk), $\delta \approx 2.5$ nm, and $d_{\text{trans}} \approx 440$ μm (used for our transport measurement), we can estimate that the bulk conductance is roughly 640 times larger than the TSS conductance in our BT sample. Thus, with such thick samples, it is hard to detect the TSS conductance using transport measurements, because it is buried within the prevailing bulk conductance. We calculated that the TSS conductance will only become comparable to the bulk conductance when the width of our BT sample is reduced to $d_{\text{trans}} \sim 0.7$ μm. For more bulk-doped samples, the thickness should be thinner. When the bulk carrier density $n_{\text{bulk}} \approx 10^{17}~\text{cm}^{-3}$, the boundary between the bulk-dominant and surface-dominant regimes forms at around $d_{\text{trans}} \approx 10$ nm [ref. 20].

In contrast, we claim that our spectroscopic mobility detection is applicable to most high-quality 3DTIs. The key to our approach is the separability of the effective masses and scattering times between the surface and bulk states. The separability can be softened by the coupling between the TSS and bulk states, mostly mediated by phonons[56,65,66]. However, the TSS-bulk coupling would not be significantly large when both states are largely separated in terms of energy and momentum (such that it is hard to interact with via phonons). In our case, the BT sample has $E_F$ at the bottom of the BCB (Fig. 2a). Therefore, it is difficult to couple the TSS and BCB because they are far away from each other in the $E$-$k$ dispersion. Therefore, for most high-quality 3DTIs, our spectroscopic acquisition of the carrier mobility will be applicable. We note that our method may be more reliable than the transport measurement in terms of separability. The phonon-assisted coupling significantly affects the magnitude of the (bulk/TSS) conductance even for high-quality 3DTIs, whereas it moderately modifies the (bulk/TSS) band dispersion and the (bulk/TSS) scattering time.

In our analyses, we assumed that the TSS mobility is mainly limited by the electron–phonon scattering. At RT, phonons usually form the dominant scattering centers, reducing the scattering time and thus carrier mobility[51,67]. There are other possible limiting factors; however, they are not prominent in our case. For example, electron–electron scattering in our BT sample should be negligible, given that the Coulomb interaction between electrons is highly screened by the strong dielectric environment (represented by the large electric permittivity $\varepsilon/\varepsilon_0 \sim 29-85$)[34]. The electron-electron scattering can diverge near the Dirac point due to the reduced screening by free carriers[68]. However, this is not suitable for our BT sample, whose $E_F$ is far (~167 meV) away from the Dirac point. Therefore, the above assumption seems to be reasonable, at least in our case.

Our observation of high TSS mobility ultimately helps to provide better consistency of TSS properties between transport and ARPES data at RT, which have remained the subject of debate[4,19,55,56,60]. This fact is crucial for resolving the issue whether or not topological quantities play a role in transport properties at RT. Also, by comparing bulk-dominant and bulk-free cases with our results, our results suggest that the role of the bulk state in hampering TSS transport properties has been overlooked, as similarly stated in ref. 69. Additionally, to estimate TSS mobility, our findings highlight the need to carefully consider the three impeding factors above, phonons, the bulk state, and impurities, together with the mobility relation.

Our findings suggest the feasibility of TI applications using the high RT mobility of the TSS, which is easily accessible without any special treatments such as gating or doping with other compounds. Therefore, the ultrahigh mobility of the TSS suggests that RT high-performance THz electronic or spintronic devices are possible with TIs[14,15]. To fully exploit the advantages of the TSS, we recommend using the THz field as both input and output signals, instead of the electrical current. Similarly, researchers have suggested using THz graphene devices to overcome the intrinsic limitation of the switching time of electronic graphene devices[70]. Furthermore, our spectroscopic method is applicable to other topological materials whose low-energy electrodynamics (e.g., the scattering time of charge carriers) provide important information. For example, long scattering times of over a few picoseconds are expected for 3D Dirac or Weyl materials[63,71]; thus, THz-TDS would be applicable for this type of an investigation, as demonstrated in this work.

**Summary**

In conclusion, our study revealed that the TSS has ultrahigh mobility even at RT. This result is surprising, given that the electron–phonon scattering limits the effective carrier transport. We measured the scattering time of the TSS carriers with THz-TDS at ~4.8 ps and obtained the effective mass of ~0.25$m_e$ from ARPES measurements. In turn, we acquired the TSS mobility value of $\mu$ (TSS) $\cong$ 33,000 cm$^2$V$^{-1}$s$^{-1}$ per TSS at RT. Among the mobilities of Bi-based 3DTIs, our TSS mobility corresponds to a record high value. Our work shows that helical Dirac fermions can be exploited directly for RT applications, which has not been achieved until now. Additionally, our approach to obtain the carrier mobility based on spectroscopic techniques can be applied to any material system, enabling synergetic collaboration of cutting-edge spectroscopy with the novel materials science community[63,71,72].

# Methods

## Laser angle-resolved photoemission spectroscopy (ARPES)

To observe the band structure of our BT crystal, we carried out ARPES and time-resolved ARPES[73], both of which show the TSS and bulk states (Fig. 2). The employed spectrometer consisted of a hemispherical analyzer and a Ti:sapphire laser system delivering 5.92-eV probe pulses at a repetition rate of 250 kHz. The fluence (the amount of incoming photon energy per area) of the probe beam was reduced until the spectral shift, due to the space charge effect, became less than 3 meV. The measurements were performed on a freshly cleaved surface of our sample, at a temperature of 300 K under quasi-vacuum conditions of $2 \times 10^{-10}$ Torr. To calibrate the Fermi level of our BT spectrum, we used the Fermi cutoff of gold in electrical contact with the sample and the analyzer as a reference. The spectral resolution was ~15 meV for our ARPES data. In our time-resolved ARPES measurements, the excitation of the sample was achieved by 800-nm irradiation (≈1.55 eV in energy) and a fluence of 160 μJ/cm$^2$.

## Terahertz time-domain spectroscopy (THz-TDS)

We performed THz spectroscopy to measure the transport properties of our BT crystal. THz-TDS enables the identification of free charge carriers residing in both the TSS and bulk states and the lattice vibrations over the frequency (time) range of 0.1–3 THz (0.33−10 ps). For the THz-TDS experiments, we prepared a freshly cleaved BT crystal (cleaved down to 6 μm) and a commercial spectrometer (TPS3000, TeraView Limited, Cambridge, UK), utilizing picosecond time-domain pulses. Sample and reference blank holder signals (Fig. 2a) were acquired through separate measurements. The measured time-domain signals were then transformed to frequency-domain THz spectra through a Fourier transformation. The calculated sample spectrum was normalized against the calculated reference spectrum, to determine the normalized transmittance of the sample (see **Supplementary S2**).

## Transfer matrix method

To acquire the optical conductivities of the TSS and bulk states in BT, we applied the transfer matrix method (TMM) to the TI model system (Fig. 3c). The TMM is typically used to analyze reflectance or transmittance of multilayer structures, in which the individual layers have different refractive indices. Given the refractive index of each layer, the optical properties of the multilayer structure were obtained by multiplying the transfer matrices of individual layers.

Let us define the following:

$$\text{Phase factor } \beta_i \equiv 2\pi \tilde{n}_i d_i \omega \ ,$$

$$\text{Reflectance } \tilde{r}_{ij} \equiv \frac{\tilde{n}_i - \tilde{n}_j}{\tilde{n}_i + \tilde{n}_j} \ ,$$

$$\text{Transmittance } \tilde{t}_{ij} \equiv \frac{2\tilde{n}_i}{\tilde{n}_i + \tilde{n}_j},$$

where i is the index of the layer counted from the incident surface, $\tilde{n}_i$ is the index of refraction of the i-th layer, and $d_i$ is the thickness of the i-th layer.

The transfer matrix of the i-th single layer is given by:

$$\text{TM}_i = \frac{1}{\tilde{t}_{ij}} \begin{pmatrix} e^{-i\beta_i} & \tilde{r}_{ij} e^{-i\beta_i} \\ \tilde{r}_{ij} e^{i\beta_i} & e^{i\beta_i} \end{pmatrix}.$$

Then, the transfer matrix of an n-layer structure is the product of the transition matrices of the n layers, as given below:

$$\text{TM}_{n-\text{layer}} = \prod_{i=1}^{n} \text{TM}_i.$$

The complex reflectance and transmittance of an n-layer system can be calculated using the following:

$$\tilde{r}_{n-\text{layer}} = \text{TM}_{n-\text{layer}}(2,1)/\text{TM}_{n-\text{layer}}(1,1);$$
$$\tilde{t}_{n-\text{layer}} = 1/\text{TM}_{n-\text{layer}}(1,1);$$

where (*x*, *y*) represents the entry of the *x*-th row and *y*-th column of the matrix. By fitting the model transmittance to our measured THz transmittance (Fig. 3c), we acquire the refractive indices of the TSS and bulk state, and therefore their optical conductivities. The analysis with TMM is applicable to other materials with nonuniform optical properties, e. g., materials with two-dimensional electronic gas or exotic interfacial states.

The results show Drude-like THz conductance spectra for the TSSs and bulk states (Fig. 3d), which exhibit distinctive mean scattering times. The calculation reveals key electronic parameters (see **Supplementary S4** and Table S1), such as the mean scattering time, DC conductivity, and plasma frequency. Also, the bulk spectra include a strong absorption at 1.7 THz, corresponding to the $E_u$-symmetric infrared-active phonon (called the *α*-phonon). This phonon has an asymmetric lineshape, due to Fano-like interference with the continuum at lower energies (see **Supplementary S4**).

**Acknowledgements**

This work was mainly supported by the Institute for Basic Science (IBS) in Republic of Korea (grant no. IBS-R009-D1 and IBS-R009-G1). M. H. Jung, J. Kim, & N. H. Jo acknowledge the financial support from the National Research Foundation of Korea (NRF) (grant no. 2017R1A2B3007918). J. H. Kim & J. Kim acknowledge the financial support from NRF (grant no. 2017R1A5A1014862) and Samsung Research Funding Center of Samsung Electronics (project no. SRFC-MA1502-01). Y.I. acknowledges the financial support from KAKENHI (grant no. 18H01148 and 17K18749) and for the sabbatical stay in IBS-CCES financially supported by the University of Tokyo. K. Kim acknowledges the financial support from NRF (grant no. 2015R1A2A1A10056200 and 2017R1A4A1015564). S. Y. Kim was supported by the Global Ph.D. Fellowship Program (grant no. NRF-2015H1A2A1034943).

**Author contributions**

B. C. P. conceived and designed the scheme. T. W. N guided and supervised the project. J. K., N. H. J., and M. H. J. prepared $Bi_2Te_3$ single crystals and performed the sample characterizations; Y. I and S. S. carried out (Tr-)ARPES with a machine in ISSP; J. K. and J. H. K. performed THz-TDS measurements with TPS3000 (TeraView) at Yonsei Univ.; B. L. and B. C. P. analyzed THz-TDS data with the help of I. K., M. C. L., K. K., J. H. K, and T. W. N; S. Y. K. performed FTIR measurements; B. L. & B. C. P. wrote the paper. All authors discussed the results and commented on the manuscript.

**Figure captions**

**Figure 1 | High-crystallinity of our Bi₂Te₃ (BT) single crystal. a,** X-ray diffraction pattern exhibiting a family of (003) peaks (red dashed lines) is in good agreement with that in the BT reference (ICSD code. 15753) without any impurity peaks. **b,** Scanning tunneling microscopy images showing the unique hexagonal structure of BT on the basal plane; the lattice constant of the *a*-axis (equivalent to the spacing between Te atoms) is 4.32 Å, consistent with the values of previous reports[74]. **c,** Transmission electron microscopy image showing well-defined unit cells (i.e., quintuple layers) along the *c*-axis without lattice deformation; the lattice constant of the *c*-axis is 30.47 Å, which is in good agreement with previous reports[74].

**Figure 2 | Angle-resolved photoemission (ARPES) spectra to identify the band structure of our BT single crystal and acquire the band mass (or carrier effective mass). a,** ARPES spectra in equilibrium. The data show the topological surface state (TSS, V-shaped line) with the Fermi level $E_F$ above 167 meV from the Dirac point ($E_D$), the Fermi wavevector of $k_F \cong 0.075$ Å$^{-1}$, and the Fermi velocity of $v_F \approx 3.4 \times 10^5$ m/s. The band mass of the TSS was $m^*_{TSS}$ ($E = E_F - E_D$) $\approx 0.25\, m_e$, where $m_e$ is the free electron mass. The spectra also show the bottom of the bulk conduction band (BCB) near $E_F$, with a weak ARPES intensity. **b,** Integrated ARPES intensity near the $\Gamma$ point ($k = 0$). The intensity was integrated within the window, over the emission angle $\theta$, $-1° \leq \theta \leq 1°$ (around $-0.01$ Å$^{-1} \leq k \leq 0.01$ Å$^{-1}$). The intensity exhibits a strong peak near $E - E_F = -0.3$ eV, originating from the TSS. In the inset, the intensity also shows the bottom of the BCB with a hump near $E - E_F = 0.1$ eV. Tails of the TSS and BCB met at the $E_F$. **c,** Time-resolved ARPES data of our BT single crystal (at 0.35 ps after photoexcitation by 1.55-eV photons); the spectra more clearly exhibit BCB dispersion (white curved lines), whose effective mass is $m^*_{BCB} \cong 0.25 m_e$. **d,** Integrated ARPES intensity around the $\Gamma$ point of nonequilibrium spectra; the intensity of the BCB is higher, whereas that of the TSS is lower.

**Figure 3 | Terahertz (THz) transmittance and THz conductivity obtained from THz-time-domain spectroscopy (THz-TDS) experiments and Drude model analysis with variational method, respectively. a and b,** The real and imaginary parts of the transmittance from experiments (squares and circles for the real and imaginary part, respectively). Experimental data were fitted with the Drude–Fano model, leading to the fitting curves (red lines). **c,** The spatial distribution of the electronic states taken into account in the fitting process. With the transfer matrix method (***Methods***), the variational optical conductivity was generated and fitted to our THz transmittance. Our BT single crystal was considered to have the TSS (blue-shaded region) nonuniformly distributed with a spatial dispersion, $\delta$,

of 2.5 nm[46-48] and bulk states (red-shaded region) uniformly distributed throughout the whole sample over 6 μm. **d**, The TSS Drude conductivity (blue-shaded region) and bulk Drude conductivity (red-shaded region) from variational method. For simplicity, the phonon spectra near 1.7 THz is not presented (for more details, see **Supplementary S4**).

**Figure 4 | Comparison of the mobility ($\mu$) of our BT sample with previously reported Bi-based topological insulators (TIs).** In our study, we separately obtained the TSS mobility per a TSS, $\mu$ (TSS) $\cong$ 33,000 cm$^2$V$^{-1}$s$^{-1}$, and bulk mobility $\mu$ (BCB) $\cong$ 1,160 cm$^2$V$^{-1}$s$^{-1}$. In terms of magnitude and separability, our RT results are highly distinct from previously reported $\mu$ values of Bi-based three-dimensional TIs[8,16-30] acquired from Shubnikov-de Hass (SdH), Hall, and other measurements at low temperature, in which the values were mainly in the range of 400–29,000 cm$^2$V$^{-1}$s$^{-1}$. (The list of references in the left-panel bar graph corresponds to the mobility, measurement temperature, and name of the mixed compounds beside the bar graph.)

**Figure 1**

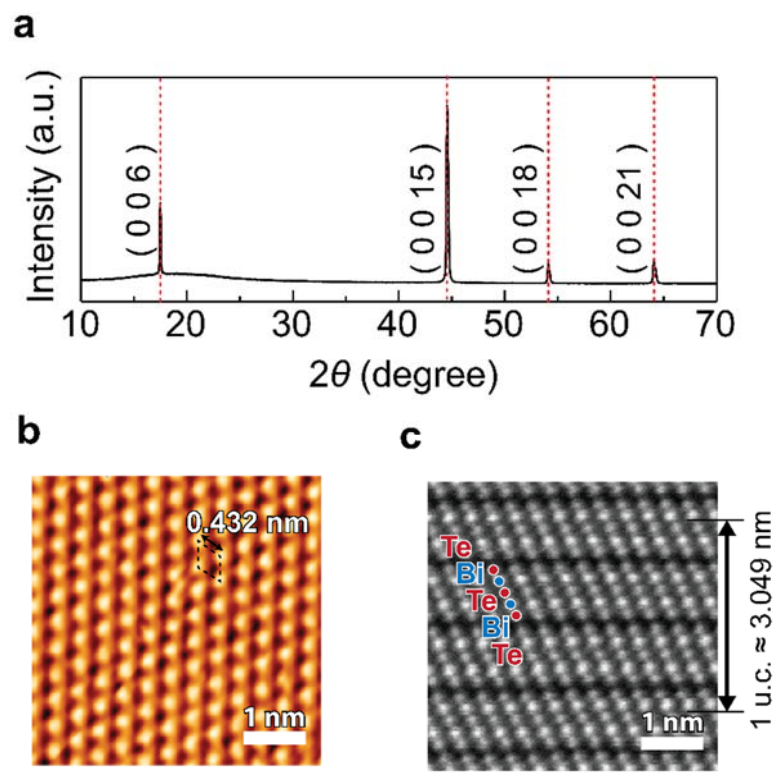

B. Lee, *et al.*

**Figure 2**

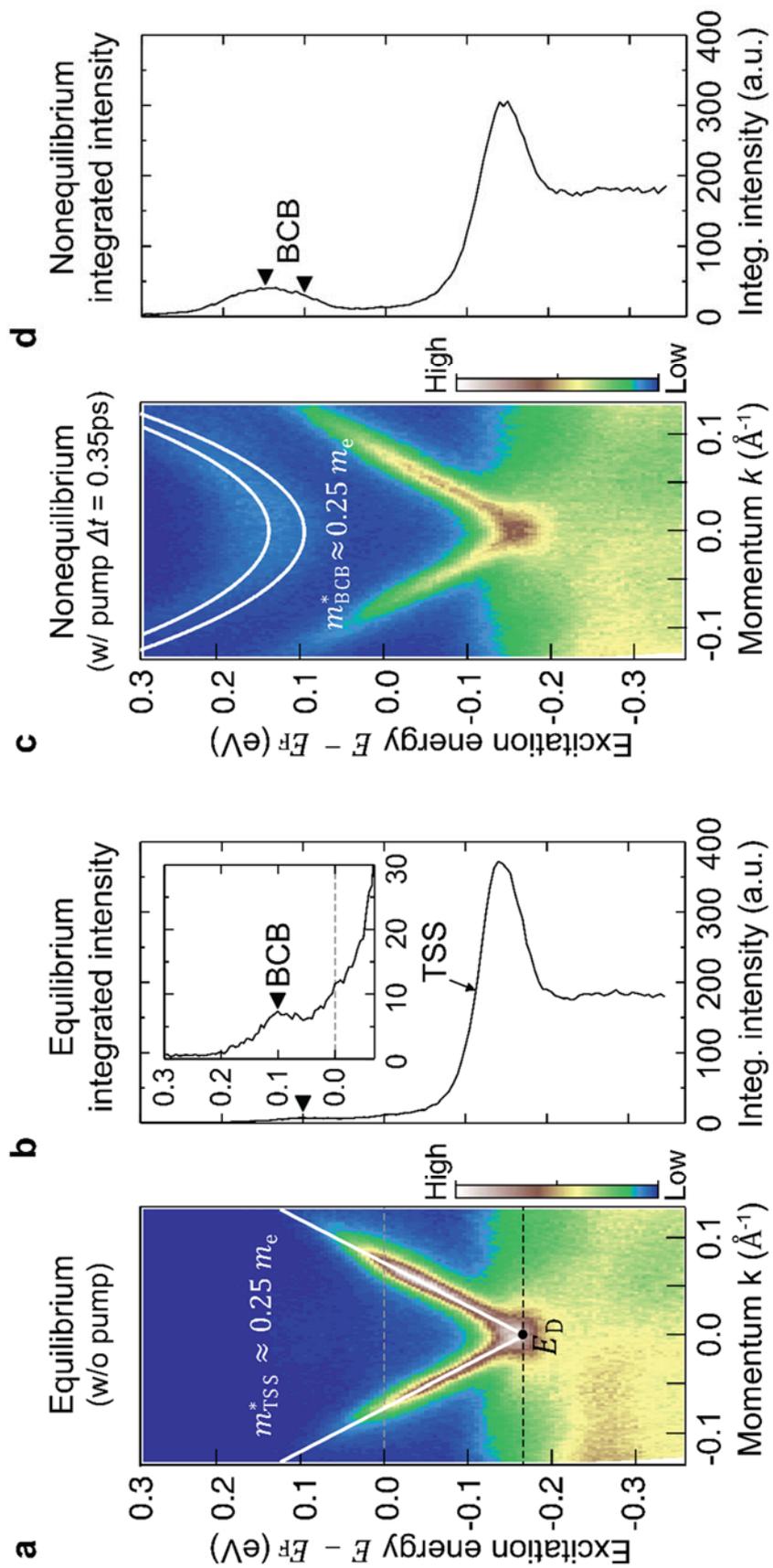

**Figure 3**

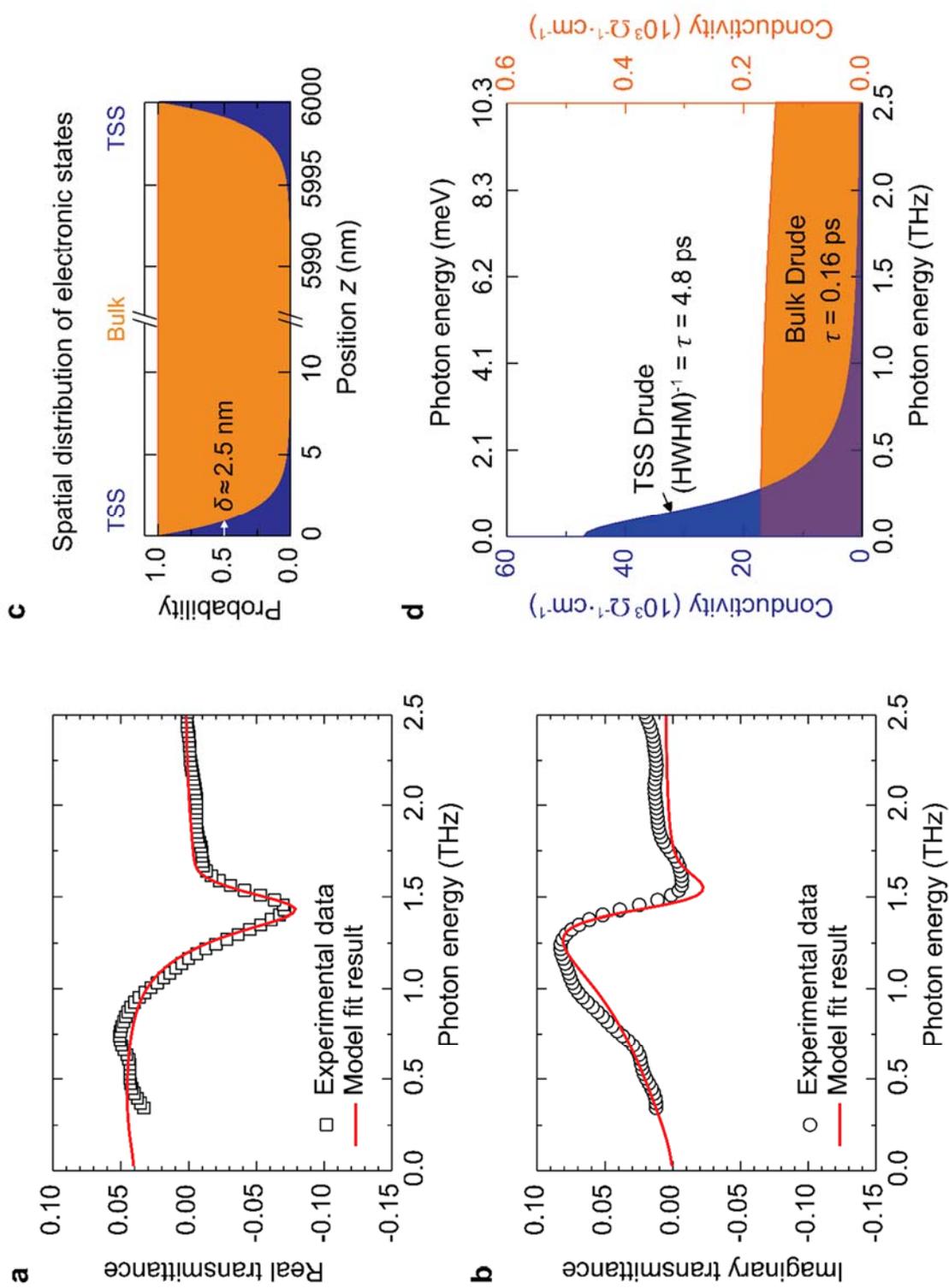

B. Lee, et al.

**Figure 4**

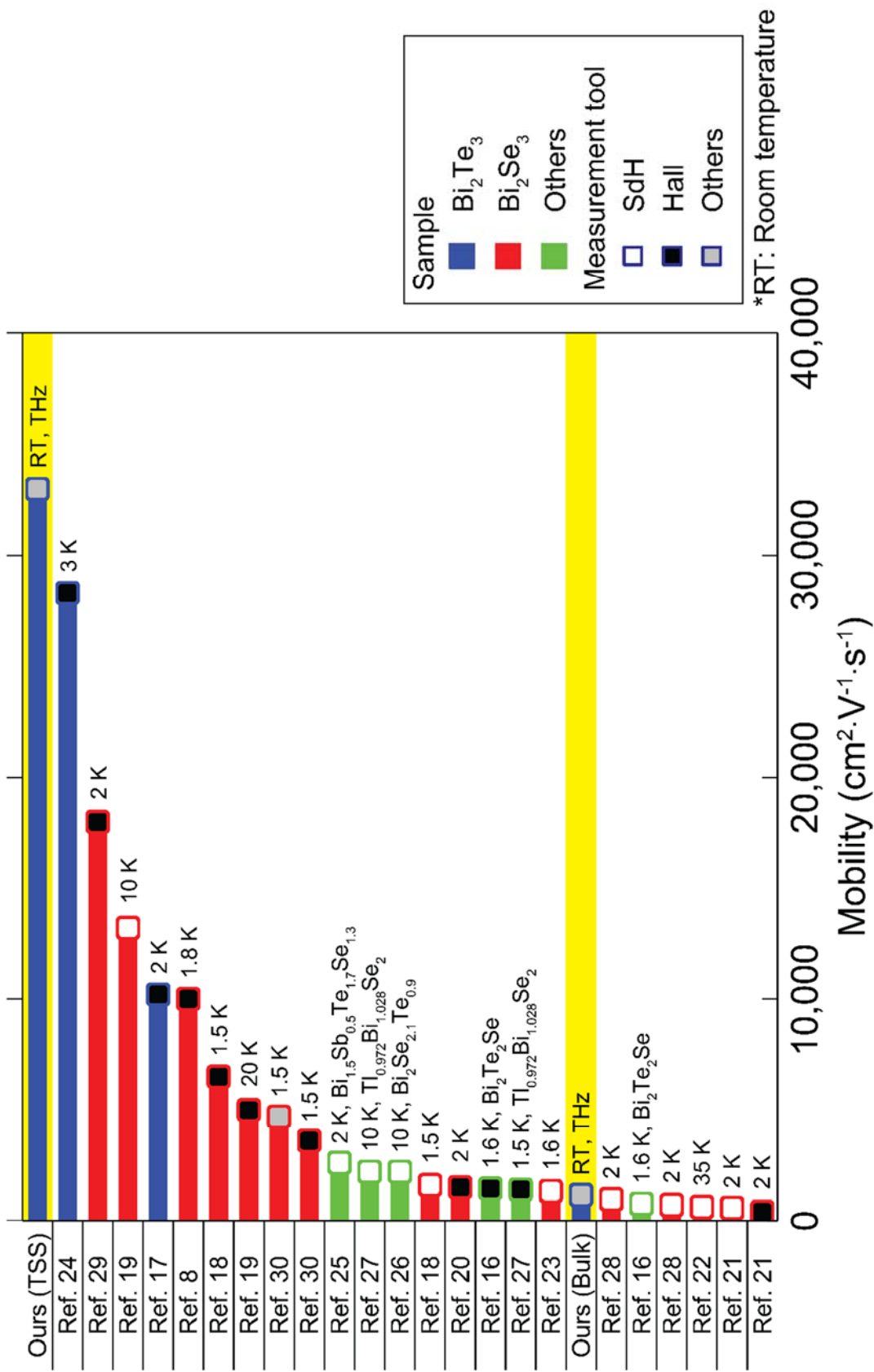

Supplementary Information for **"Revealing electrically undetectable room temperature surface-mobility of bulky topological insulators by spectroscopic techniques"**

**Authors**

Bumjoo Lee[1,2,*], Jinsu Kim[3], Jonghyeon Kim[4], Na Hyun Jo[3,#], Yukiaki Ishida[5], So Yeun Kim[1,2], Min-Cheol Lee[1,2], Inho Kwak[1,2], Shik Shin[5], Kyungwan Kim[6], Jae Hoon Kim[4], Myung-Hwa Jung[3], Tae Won Noh[1,2], and Byung Cheol Park[1,2,†]

**Affiliation**

[1]Center for Correlated Electron Systems, Institute for Basic Science (IBS), Seoul 08826, Republic of Korea

[2]Department of Physics and Astronomy, Seoul National University (SNU), Seoul 08826, Republic of Korea

[3]Department of Physics, Sogang University, Seoul 04107, Republic of Korea

[4]Department of Physics, Yonsei University, Seoul 03722, Republic of Korea

[5]ISSP, University of Tokyo, Kashiwa, Chiba 277-8581, Japan

[6]Department of Physics, Chungbuk National University, Cheongju 28644, Republic of Korea

*First author (B. Lee, kimphysics@snu.ac.kr)

†Corresponding author (B. C. P, topologicalmeta@gmail.com)

#Current affiliation (N. H. Jo): Ames Laboratory, Iowa State University, Ames, Iowa 50011, USA

## Supplementary S1. Electrical resistivity of our Bi$_2$Te$_3$ single crystal

First, in this section, we want to confirm that our sample is nearly bulk-insulating based on electrical measurements. In the main text, we referred to angle-resolved photoemission spectroscopy (ARPES) and terahertz (THz) data measured by optical techniques to show that our Bi$_2$Te$_3$ (BT) sample was nearly bulk-insulating. Here we present electrical resistivity results as a function of temperature, as shown in Fig. S1. The oscillatory behavior of the resistivity supports that the Fermi level of our BT sample lies at or below the conduction band bottom, similar to the previous result[1]. The unexpected bulk response in our results, therefore, mainly originates from thermal activation at room temperature (RT), which is apparent up to ~100 meV [Ref. 2].

Second, from the resistivity data, we wanted to emphasize that the topological surface state (TSS) and bulk responses cannot be separated at RT; this also holds true at low temperatures with electrical measurements. Particularly, the spectroscopic methods applied in this study are capable of separating both responses completely, even at RT.

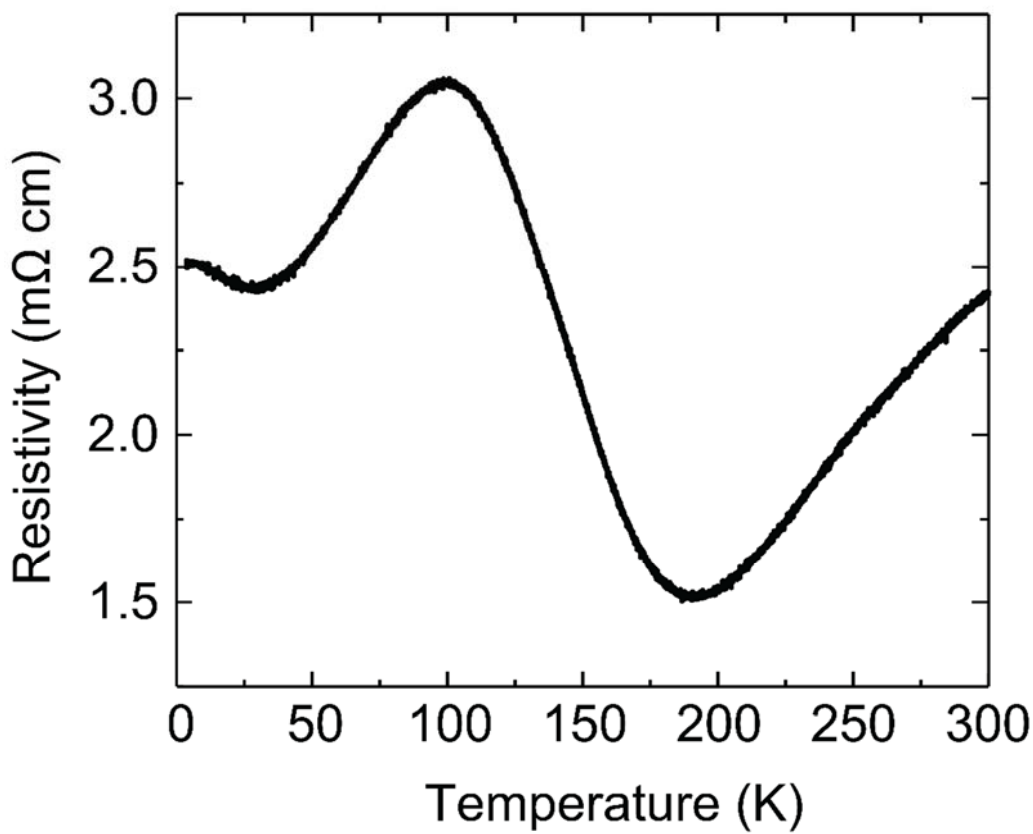

**Figure S1 | Temperature-dependent resistivity of our Bi$_2$Te$_3$ (BT) crystal.** The oscillatory behavior provides evidence of a nearly bulk-free sample[1]. From the data, the topological surface state (TSS) and bulk responses cannot be separated, in contrast to our spectroscopic method.

**Supplementary S2. Terahertz transmittance of our Bi$_2$Te$_3$ single crystal**

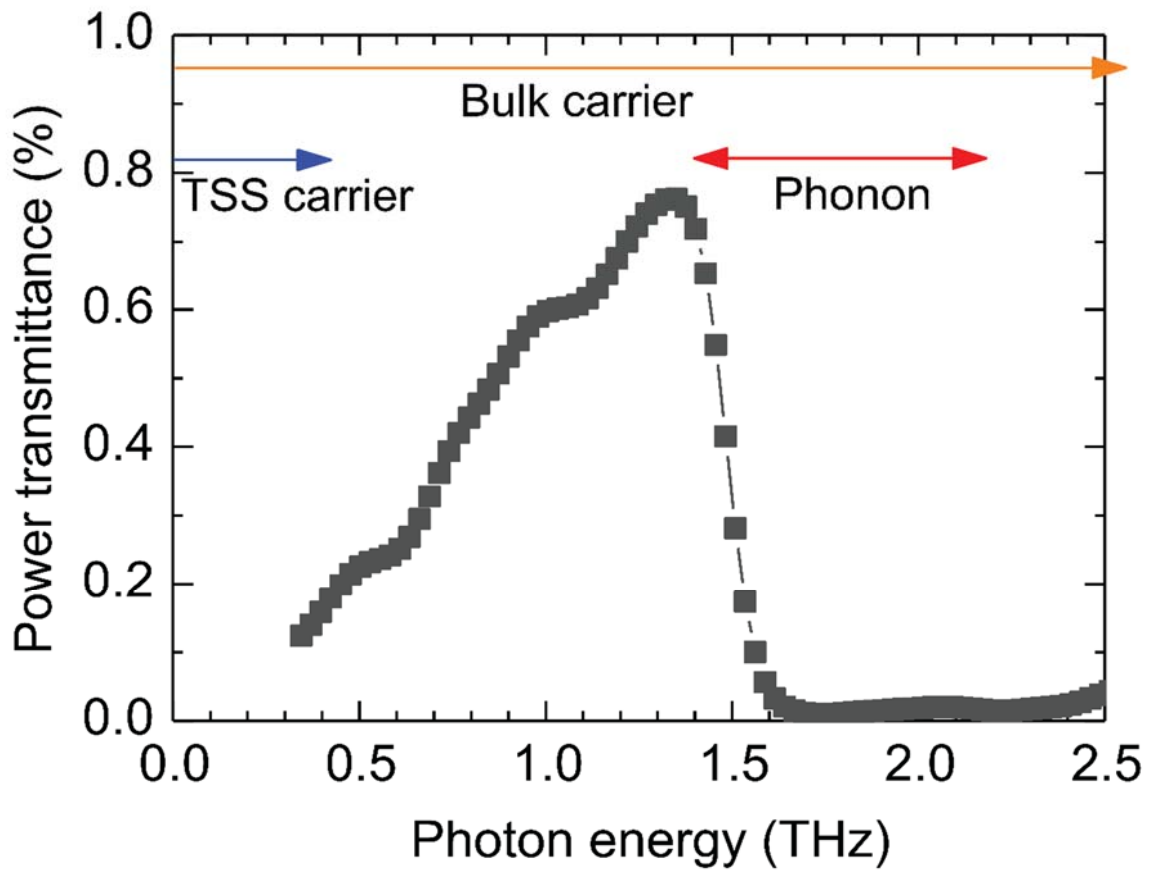

**Figure S2 | Terahertz (THz) power transmittance of our BT crystal.** The complex transmittance was obtained through Fourier transformations of reference and sample time-domain signals: $\tilde{t}(\omega) = E_{\text{sample}}(\omega)/E_{\text{ref}}(\omega)$. The power transmittance $|\tilde{t}(\omega)|^2$ over 0.3–2.5 THz (1 THz $\cong$ 4.1 meV) is shown. $|\tilde{t}(\omega)|^2$ within 0.3−1.5 THz shows relatively large values of over 0.1–0.8%, which guarantee the reliability of further analysis. $|\tilde{t}(\omega)|^2$ reflects the TSS (blue arrow) and bulk (orange arrow) Drude peaks[3] and phonon peak (red arrow) near 1.7 THz[4,5]. $|\tilde{t}(\omega)|^2$ shows the one-to-one correspondence to conductivity spectra in Fig. 3d in the main text (arrows: twice the width of the peak).

## Supplementary S3. Plasma frequency of Dirac fermions

S3-1. Fourier transform infrared spectroscopy (FTIR) measurements

To determine the plasma frequency of the Dirac fermions, we performed FTIR reflectance measurements on our BT sample (Fig. S3) by a commercial spectrometer (Bruker Vertex 80v). A gold normalization technique was used to obtain the correct values of reflectance[6]. For metallic samples, FTIR spectra showed a reflectance threshold (referred to as the plasma edge) corresponding to the (screened) plasma frequency of charge carriers[7].

If we consider both bulk and TSS Drude responses with (bare) plasma frequencies ($\omega_{p,Bulk}$ and $\omega_{p,TSS}$) and mean scattering times ($\tau_{Bulk}$ and $\tau_{TSS}$), the dielectric function can be expressed as follows:

$$\varepsilon = \varepsilon_\infty - \left(\frac{\omega_{p,Bulk}^2}{\omega + i/\tau_{Bulk}} + \frac{\omega_{p,TSS}^2}{\omega + i/\tau_{TSS}}\right)\frac{1}{\omega} , \qquad \text{Eq. (S1)}$$

where $\varepsilon_\infty \cong 64$ is the high-frequency limit of the dielectric function corresponding to the screening of bound electrons[8]. Suppose that $1/\tau_{(Bulk,TSS)} \ll \omega_{p,(Bulk,TSS)}$. Then the reflectance threshold occurs around $\omega_{RT} \cong \sqrt{\frac{\omega_{p,Bulk}^2 + \omega_{p,TSS}^2}{\varepsilon_\infty}}$. Furthermore, $\omega_{RT}$ can be approximated as $\omega_{RT} \cong \sqrt{\frac{\omega_{p,TSS}^2}{\varepsilon_\infty}} = \omega_{p,TSS}^*$, due to the stronger metallicity of the TSS. Finally, we obtained the bare plasma frequency of the TSS carriers, $\omega_{p,TSS}$ (FTIR) $\cong$ 544 meV.

S3-2. ARPES measurements

From the ARPES data (Fig. 2a in the main text), we directly obtained the Fermi wavevector, $k_{F,ARPES} = 0.075$ Å$^{-1}$. Considering a two-dimensional free electron gas, the sheet carrier density of a single TSS[9] is $n_{TSS}$ (exp) $= k_{F,ARPES}^2/4\pi \cong 4.4 \times 10^{12}$/cm$^2$. To calculate the plasma frequency of the TSS, $\omega_{p,TSS}$ (ARPES), from ARPES data, we used the equation giving the Drude weight of the TSS[10-12]

$$D_{TSS} = \pi\,(\omega_{p,TSS}^2 \cdot \delta)/60 = (v_F e^2/4\hbar)\sqrt{\pi|n_{TSS}|} , \qquad \text{Eq. (S2)}$$

where $\omega_{p,TSS}$ is the plasma frequency, $\delta \cong 2.5$ nm is the penetration depth of the TSS wavefunction (Fig. 3c)[13-15], $v_F$ is the Fermi velocity, $e$ is the elementary charge, and $\hbar$ is the reduced Planck constant. Note that the topological insulator (TI) possesses no spin or valley degeneracies, in contrast to graphene in which the Drude weight has four-fold degeneracies. As a result, we found $\omega_{p,TSS}$ (ARPES) $= 550$ meV, which is consistent with the FTIR result above, $\omega_{p,TSS}$ (FTIR) $\cong$ 544 meV.

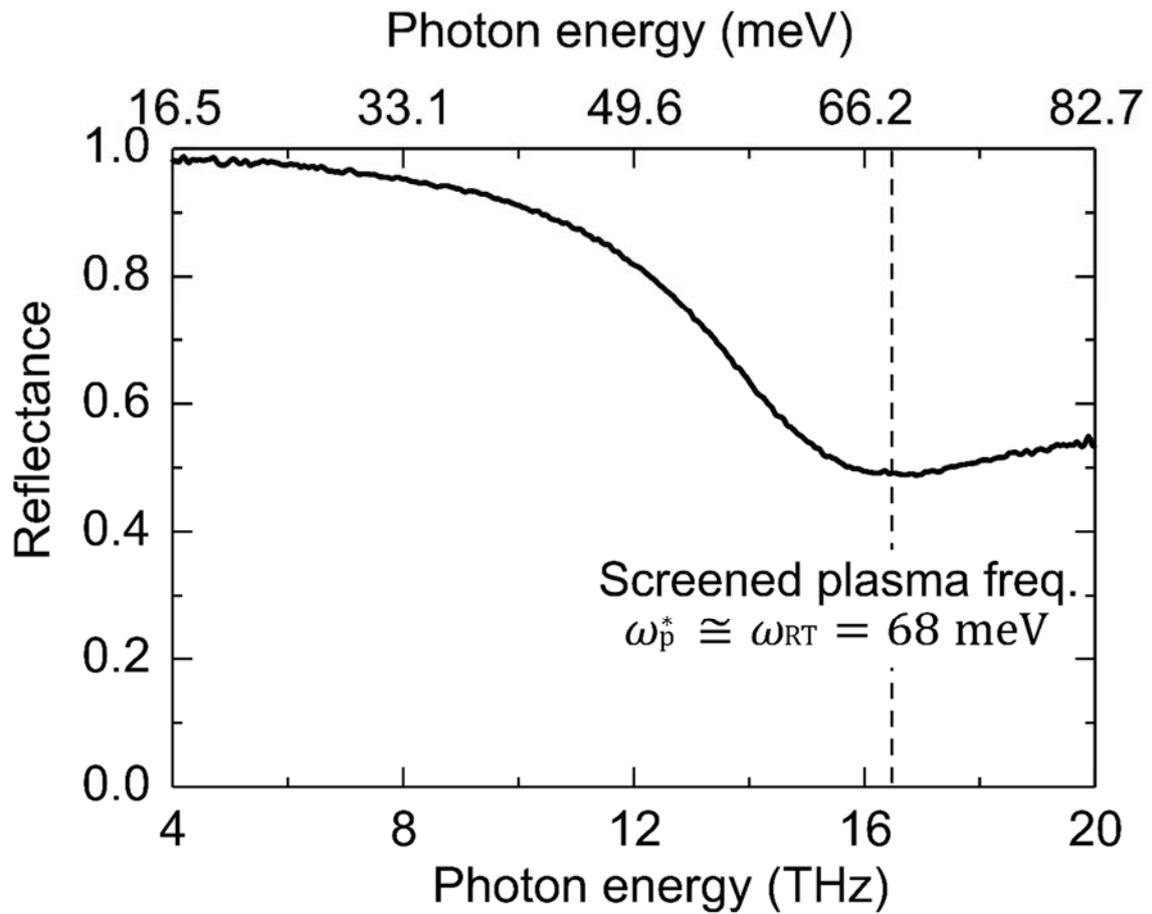

**Figure S3 | Far-infrared (FIR) reflectance spectra of our BT crystal.** The spectra provide the reflection threshold (i.e., screened plasma frequency), $\omega_p^* \cong \omega_{RT} = 68 \text{ meV}$, which was used to determine the TSS Drude peak.

## Supplementary S4. Conductivity model and fitting results

S4-1. Optical conductivity $\sigma(\omega)$ model

In TIs, there are two types of charge carriers originating from the TSS and bulk state, as described in Fig. 3c in the main text. They exhibit free carrier responses, resulting in two Drude components. In addition, the infrared-active $E_u^1$ phonon appears over the bulk.

Therefore, we set the $\sigma(\omega)$ model for the TSS below:

$$\sigma_{\text{TSS}}(\omega) = \frac{\pi}{15}\frac{1}{4\pi}\frac{\tau_{\text{TSS}}\omega_{p,\text{TSS}}^2}{1-i\omega\tau_{\text{TSS}}}\ ,\qquad \text{Eq. (S3)}$$

where $\omega_{p,\text{TSS}}$ is the plasma frequency and $\tau_{\text{TSS}}$ is the scattering time of the TSS carriers. The $\sigma(\omega)$ model for the bulk is described by the following[5,16]:

$$\sigma_{\text{Bulk}}(\omega) = \frac{\pi}{15}\frac{1}{4\pi}\frac{\tau_{\text{bulk}}\omega_{p,\text{bulk}}^2}{1-i\omega\tau_{\text{bulk}}} + \frac{\pi}{15}\frac{i}{4\pi}\frac{\omega\Omega_{p,\text{ph}}^2}{(\omega^2-\omega_0^2)+i\omega\Gamma_{\text{ph}}}\frac{(q-i)^2}{|q^2-1|}\ ,\qquad \text{Eq. (S4)}$$

where $\omega_{p,\text{bulk}}$ and $\tau_{\text{bulk}}$ are the plasma frequency and scattering time of the bulk carriers, respectively, and $\Omega_{p,\text{ph}}$, $\omega_0$, $\Gamma_{\text{ph}}$, and $q$ are the oscillator strength, center frequency, linewidth, and asymmetry parameter of the phonons, respectively. Note that we used an asymmetric Lorentz model (or Fano model)[16] to describe the phonon whose asymmetry arises from the quantum interference of phonons with the electronic continuum[17]. Here, a larger $q$ indicates a more symmetric (Lorentzian-like) response.

S4-2. Drude–Fano model fit results

Based on the above model, we constructed a transmittance model (refer to **Methods** in the main text) to compare with the transmittance obtained experimentally. The best fitting results are shown in Fig. S4a-d, giving a minimal fit residual. To enhance fitting reliability, we adopted physical quantities from FTIR and ARPES measurements, including the plasma frequency $\omega_p$ and carrier density $n$ (see **Supplementary S3**). The plasma frequency squared of TSS carriers is roughly nine-fold greater than that of bulk carriers, which indicates the stronger metallicity of TSS. Our fitting results are consistent with experimental data (shown in Fig. 3a and b in the main text). The resulting (real/imaginary) optical conductivity is presented in Fig. S4(a,b/c,d) below. Note that we assigned the sharp Drude component to the TSS response and the broad Drude component to the bulk response (The assignment is justified in detail in **Supplementary S6** and consistent with Park *et al.*[5].).

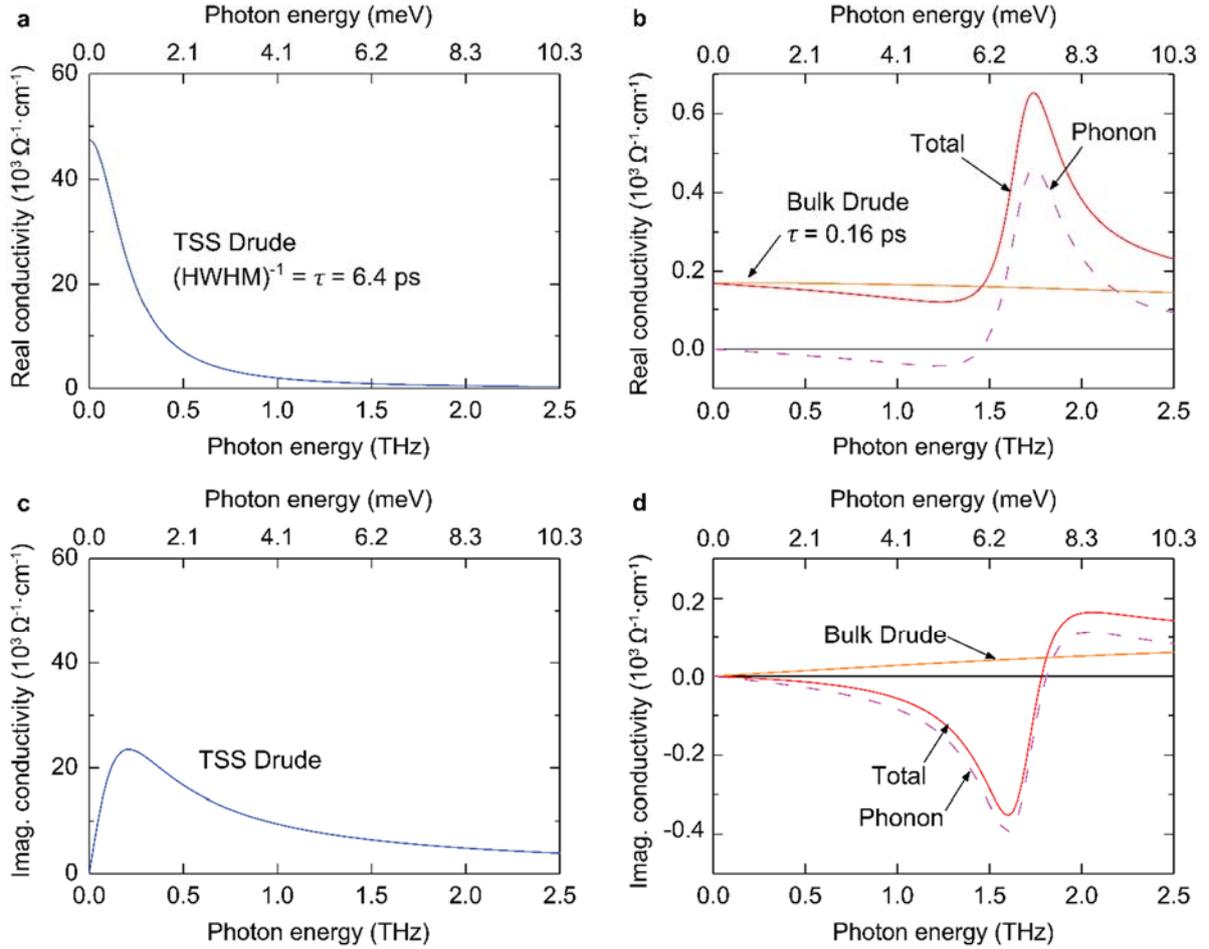

**Figure S4 | THz conductivity spectra acquired from the Drude–Fano model fit. a**, The resulting TSS Drude real conductivity (blue line). **b**, The total bulk real conductivity (red line), comprised of the bulk Drude conductivity (orange line) and phonon conductivity (purple dashed line). **c**, The resulting TSS Drude imaginary conductivity (blue line). **d**, The total bulk imaginary conductivity (red line), comprised of the bulk Drude conductivity (orange line) and phonon conductivity (purple dashed line). All of the corresponding fitting parameters are displayed in table S1.

**Table S1 | Fit parameters for the Drude–Fano model.** Note that the $\omega_p$ values are the bare plasma frequencies of the charge carriers calculated with the high-frequency limit of the dielectric constant $\varepsilon_\infty$ = 64 [Ref. 8]. For the TSS, the displayed fitting parameters can be determined by a constraint; the plasma frequency of the nonuniform TSS was fixed to the value uniquely acquired from uniform model calculations.

|  | TSS Drude | Bulk Drude |  | Phonon |
|---|---|---|---|---|
| **Plasma frequency $\omega_p$** | 132.96 THz (549.88 meV) | 44.06 THz (177.41 meV) | **Oscillator strength $\Omega_{p,ph}$** | 16.10 THz (66.55 meV) |
| **Mean scattering time $\tau$** | 4.76 ps | 1.67 ps | **Phonon linewidth $\Gamma_{ph}$** | 0.34 THz (1.41 meV) |
|  |  |  | **Center frequency $\omega_0$** | 1.69 THz (6.99 meV) |
|  |  |  | **Fano parameter $q$** | 3.29 |

**Supplementary S5. Appropriateness of a nonuniform model fit: comparison with a uniform model fit**

Conventionally, the optical conductivity of a single crystalline sample can be obtained through Kramers-Kronig relations[18]. If the optical properties of a sample are uniform throughout the beam path, the optical conductivity of the sample can be obtained unambiguously with algebraic method (uniform model). However, in the main text, we used a variational method with a nonuniform model (Fig. 3c) for acquiring the optical conductivity. Here, we justify the goodness of the nonuniform model by comparing its fitting quality with that of a uniform model. Figure S5 shows a comparison between uniform (black line) and nonuniform (red line) models; the nonuniform model better represents the experimental data (symbols), showing a more consistent fit than the uniform model.

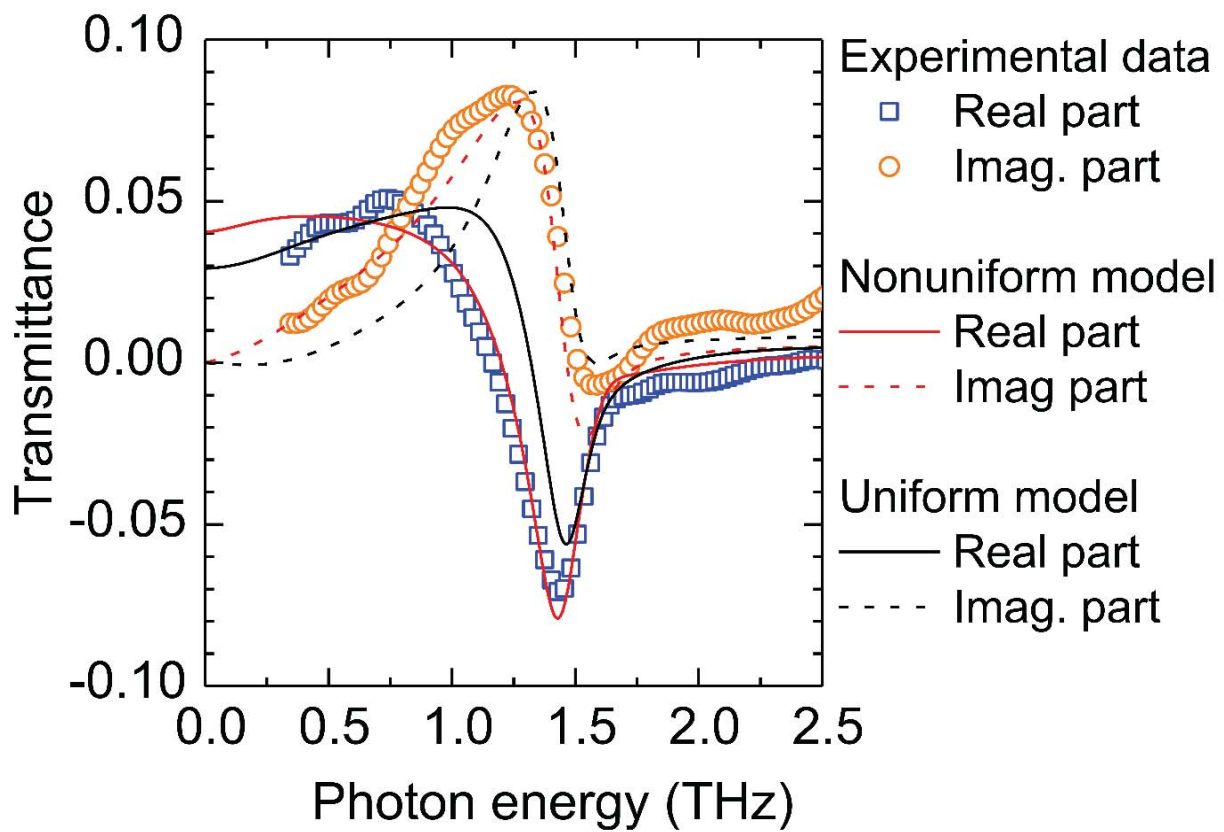

**Figure S5 | Experimental transmittance (symbols) fitted with the model transmittance from the nonuniform model and calculated with the model optical conductivity from the uniform model.** The nonuniform model produces the complex model transmittance coincident with the experimental transmittance better than the uniform model.

## Supplementary S6. Assignment of Drude components: sharp (broad) Drude peak as the TSS (bulk) carrier response

In this section, we are going to assign the Drude components to the electronic state from which they originated. There are two possible cases: Case I and Case II, as described in Fig. S6a and S6b, respectively. For simplicity, in this specific case, we employ the uniform conductivity model instead of the nonuniform conductivity model. Note that this model is sufficient for assigning the Drude peaks, as described below.

For the assignment, we compare the TSS carrier densities calculated for cases I and II with the experimental value, $n_{\text{TSS}}(\exp) \cong 4.4 \times 10^{12} \text{cm}^{-2}$ (see **Supplementary S3**). Case I provides $n_{\text{TSS}}(\text{Case I}) \cong 1.1 \times 10^{14} \text{cm}^{-2}$, which is 25 times greater than the experimental value, whereas Case II gives $n_{\text{TSS}}(\text{Case II}) \cong 5.4 \times 10^{16} \text{cm}^{-2}$, about 12,000 times greater than the experimental value. Therefore, we can conclude that Case I is more realistic, despite some deviation. However, we emphasize that using a uniform model to represent a nonuniform medium, such as that of a TI, can lead to such deviations. In fact, Case I conforms our intuition. The TSS exhibited a helical spin texture, suppressing the backscattering of charge carriers from impurities. Therefore, the scattering rate of the TSS carriers is naturally less than that of the bulk carriers. Although electron–phonon scattering occurs at RT, electron–impurity scattering remains as the key factor for distinguishing the two Drude responses, as discussed in the main text.

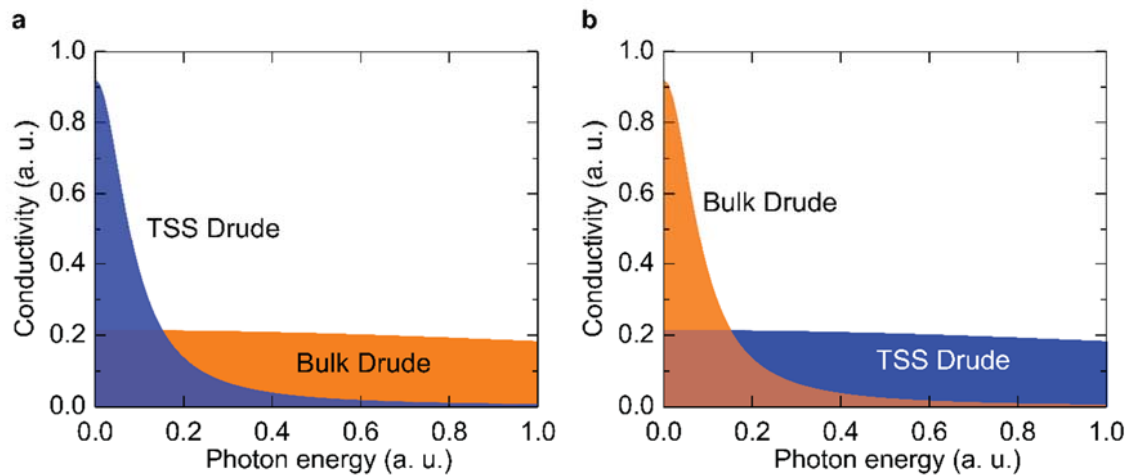

**Figure S6 | Schematic illustration of two cases for the Drude assignment. a** (Case I), a sharp Drude component with a low (long) scattering (rate) time is assumed to be the topological surface state (TSS) response. **b** (Case II), a broad Drude component with high (short) scattering rate (time) is assumed to be the TSS response. We found that Case I is the proper case.

## Supplementary S7. Terahertz signature of helical Dirac fermions

Raw THz data clearly exhibited the signature of helical Dirac fermions residing in the TSS. This is in sharp contrast to conventional transport data, in which both the TSS and bulk signals are mixed. In the THz data, the helical Dirac fermions highly absorb the low-energy THz signal, leading to strong absorption below 3 meV. This appears as a downturn in the real part of the transmittance and dielectric function, as shown in Fig. S7.

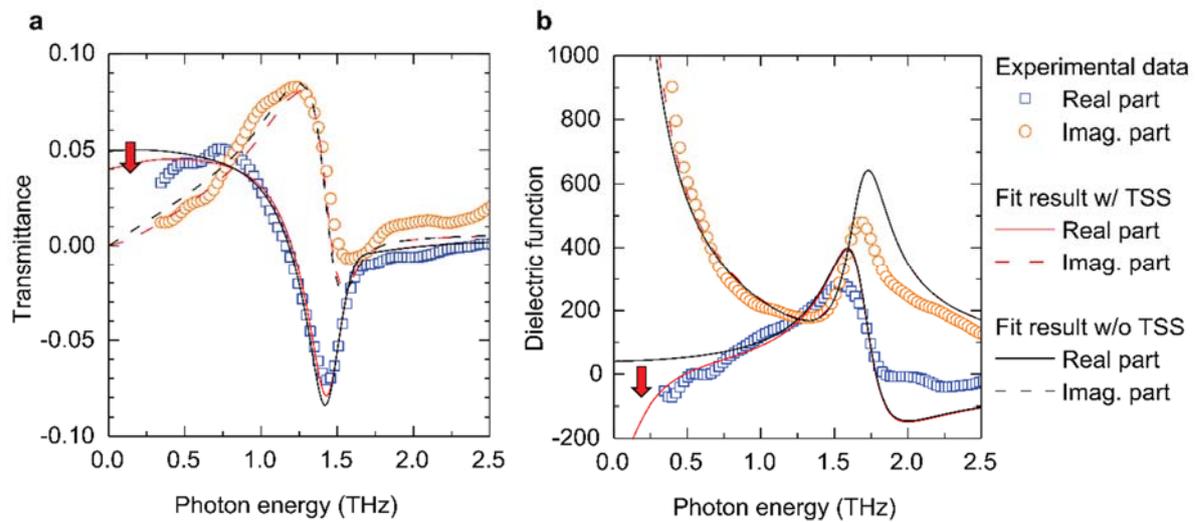

**Figure S7 | THz signature of the helical Dirac fermions, with strong absorption below 3 meV. a,** Experimental real (rectangles) and imaginary (circles) parts of the transmittance are plotted. Model complex transmittance with (red lines) and without (black lines) involvement of the response of helical Dirac fermions are also presented. **b,** Calculated real (rectangles) and imaginary (circles) parts of the dielectric function based on the uniform model. Model complex dielectric functions, with (red lines) and without (black lines) involvement of the response of helical Dirac fermions, are also presented.

## Supplementary S8. Precision of the TSS scattering time

The signature of the TSS is the downturn of the real transmittance below 0.5 THz (see **Supplementary S7**). Therefore, fitting in the low-frequency region is important for extracting the scattering rate of the TSS carriers. To capture the signature of TSS precisely, weighting the low-frequency region is advantageous.

Our approach to weighting the low frequency involved first defining the error function, $\sum_\omega t_{exp}(\omega) - t_{TMM}(\omega)$, where $t_{exp}(\omega)$ is the experimentally measured transmittance and $t_{TMM}(\omega)$ is transmittance calculated with the transfer matrix method (TMM) (see **Methods**). Then, the error function is multiplied by $1/\omega^n$, where *n* is the weight power.

As shown in Fig. S8a, the scattering rate changes as the weight power varies. Specifically, the scattering time overall increased within $0.05 \leq n \leq 2$. The scattering time is undefined at both ends (gray shaded region) where the fit did not converge.

To obtain the ideal *n*, we calculated the fit residual (defined by the average of $|t_{exp}(\omega) - t_{TMM}(\omega)|/|t_{exp}(\omega)|$) over the entire range from 0–6.2 meV (black), including all spectral components and that below 3.2 meV (red), where the TSS Drude peak is dominant (Fig. S8b). To acquire the precise scattering rate of the TSS carriers, *n* = 1 was found to provide the ideal solution, in which both residuals were minimized.

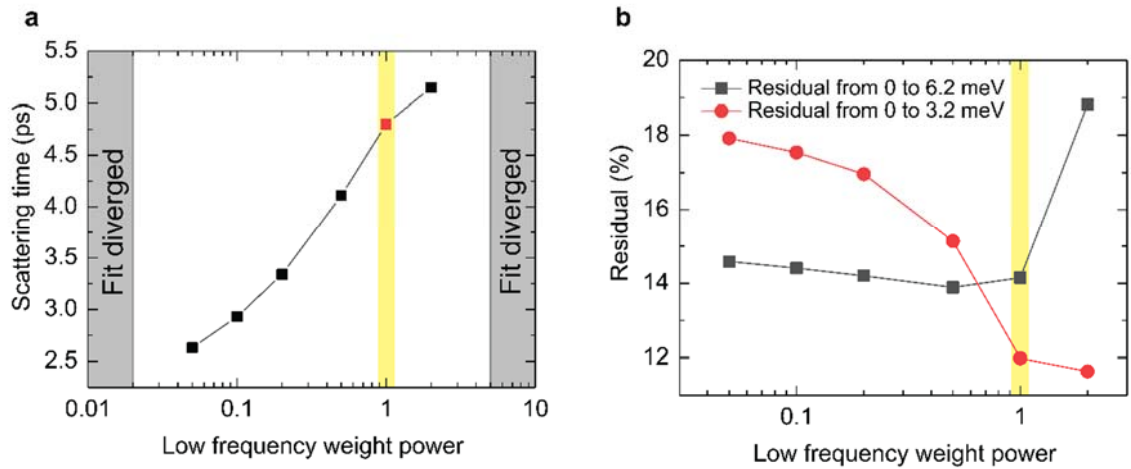

**Figure S8 | Fitting parameter and fit residual change by low-frequency fitting weight power. a,** The scattering time of the TSS carriers relying on the low-frequency weight power *n*. **b,** The fit residuals (defined by the average of $|t_{exp}(\omega) - t_{TMM}(\omega)|/|t_{exp}(\omega)|$) over 0–6.2 meV (red) and 0–3.2 meV (black), relying on *n*. The results indicate that *n* = 1 is the most optimal choice, providing minimal fit residuals.

**Supplementary S9. Comparison of our bulk Drude response to the free carrier response of TIs in the literature**

Interestingly, our bulk Drude peak was similar, in terms of the scattering time of the free carrier response (including the TSS and/or bulk carriers), to those of TIs in the literature[4,18,19]. To show their similarity directly, we plotted the Drude responses from the literature together with our bulk Drude peak (Fig. S9). The spectra are normalized by the (extrapolated) DC conductivity for comparison.

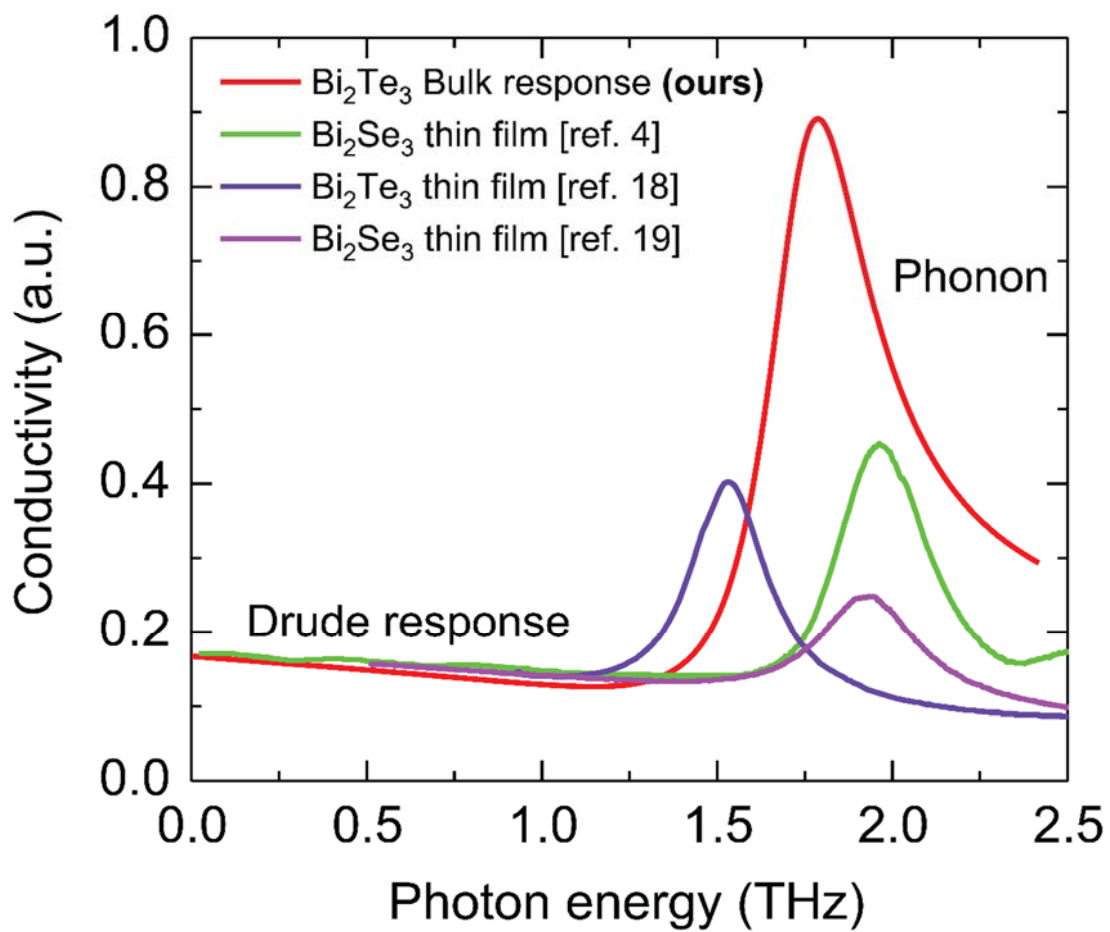

**Figure S9 | Comparison of our bulk Drude response to the Drude response of topological insulators (TIs) from the literature[4,18,19].** The responses of the charge carriers and phonons are observed at around 1.5–2.0 THz. Only a single Drude peak appears in the literature, and it is similar to the bulk Drude peak in our data. Note that the mismatch of the phonon frequency may come from the difference in the constituent element (Se/Te) and the sample geometry (fil)m/single crystal).